\documentclass[lettersize,journal]{IEEEtran}
\usepackage{amsmath,amsfonts}
\usepackage{algorithmic}
\usepackage{algorithm}
\usepackage{array}
\usepackage[caption=false,font=normalsize,labelfont=sf,textfont=sf]{subfig}
\usepackage{textcomp}
\usepackage{stfloats}
\usepackage{url}
\usepackage{verbatim}
\usepackage{graphicx}
\usepackage{cite}
\usepackage[cp1252]{inputenc}
\usepackage{xcolor}
\begin{document}

\title{Smart Handover with Predicted User Behavior using Convolutional Neural Networks for WiGig Systems} 

\author{Tiago~Koketsu~Rodrigues,~\IEEEmembership{Member,~IEEE,}
		Shikhar~Verma,~\IEEEmembership{Member,~IEEE,}
		Yuichi~Kawamoto,~\IEEEmembership{Member,~IEEE,}
	       Nei Kato,~\IEEEmembership{Fellow,~IEEE,}
	       Mostafa M. Fouda,~\IEEEmembership{Senior Member,~IEEE,}
	       and~Muhammad Ismail,~\IEEEmembership{Senior Member,~IEEE}
\thanks{T. K. Rodrigues, S. Verma, Y. Kawamoto, and N. Kato are with the Graduate School of Information Sciences, Tohoku University, Sendai, Miyagi, 980-8579, Japan (e-mail: tiago.gama.rodrigues, shikhar.verma, youpsan, kato@it.is.tohoku.ac.jp).}
\thanks{M. M. Fouda is with the Department of Electrical and Computer Engineering, College of Science and Engineering, Idaho State University, Pocatello, ID 83209, USA (e-mail: mfouda@ieee.org).}
\thanks{M. Ismail is with the Department of Computer Science, College of Engineering, Tennessee Tech University, Cookeville, TN 38506, USA (e-mail: mismail@tntech.edu).}}



\maketitle

\begin{abstract}
WiGig networks and 60 GHz frequency communications have a lot of potential for commercial and personal use. They can offer extremely high transmission rates but at the cost of low range and penetration. Due to these issues, WiGig systems are unstable and need to rely on frequent handovers to maintain high-quality connections. However, this solution is problematic as it forces users into bad connections and downtime before they are switched to a better access point. In this work, we use Machine Learning to identify patterns in user behaviors and predict user actions. This prediction is used to do proactive handovers, switching users to access points with better future transmission rates and a more stable environment based on the future state of the user. Results show that not only the proposal is effective at predicting channel data, but the use of such predictions improves system performance and avoids unnecessary handovers.
\end{abstract}


\section{Introduction} 

\begin{figure}[!t]
\centering
\includegraphics[width=\linewidth]{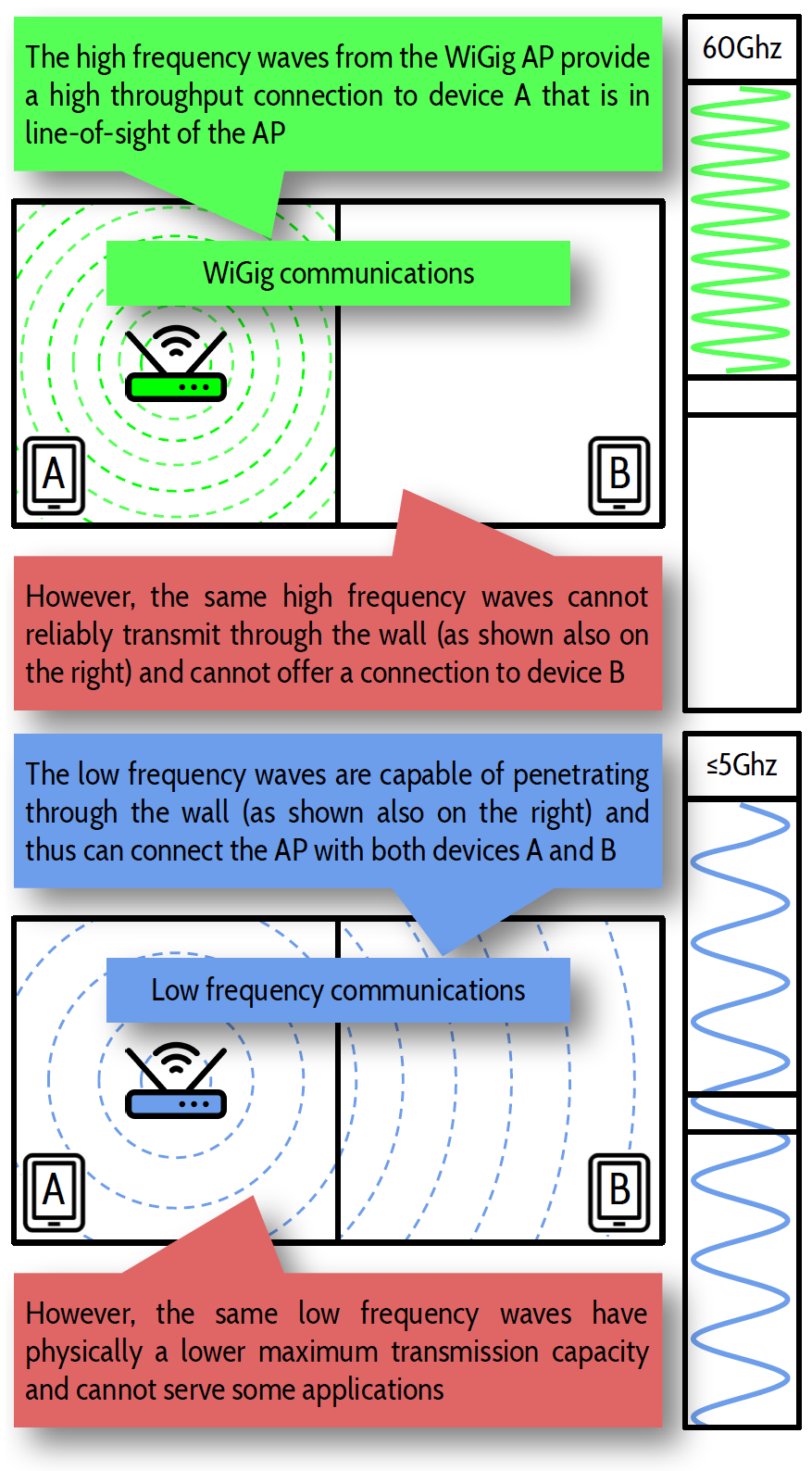}
\caption{Simplified comparison between WiGig and low frequency communications, and the tradeoff between speed and coverage.}
\label{wigig_comparison}
\end{figure}

\IEEEPARstart{W}{iGig} has been touted as the new revolutionary standard for WiFi since at least the announcement of protocol IEEE 802.11ad in 2009. Its main benefits stem from operating in the 60GHz spectrum. The higher frequency bands allow it to provide transmission rates in the range of multiple gigabits per second. These higher transmission rates can be important to support, for example, 6G and its new applications \cite{wang_quantumSmartNetworking, guo_6Gapplications}. However, despite the official addition of standard 802.11ad in 2012, WiGig never really took off in popularity or usage. The main drawbacks also come from the use of high-frequency communications. Signals in those bands have high dissipation rates, causing their range to be significantly shorter than more commonly used 2.4GHz and 5GHz bands, and extremely poor penetration rates, being almost entirely negated by most obstacles \cite{song_wigigComm}. This comparison is illustrated in \figurename~\ref{wigig_comparison}, where in the scenario on the left, although $A$ receives a high-speed connection, $B$ cannot be reached due to the wall separating it from the access point (AP). In comparison, in the scenario on the right, both devices are able to connect, but now $A$ has \IEEEpubidadjcol a lower connection speed. Such issues limit 802.11ad APs to single, small rooms where line-of-sight is guaranteed and communication distance is short \cite{bi_highFrequencyComm}. \par

In an attempt to improve on the shortcomings of 802.11ad, IEEE 802.11ay was released in 2021. Compared to its predecessor, 802.11ay introduces channel bonding, MIMO capability, and higher modulation schemes. These changes not only increase the transmission rate (from a maximum of 7Gbps to 40Gbps) but also the range of communication. With the application of low loss and high output transmitters that are already available in the market, WiGig, in the form of 802.11ay, has proven that it can deliver wireless gigabit communications even with ranges of 500 meters \cite{fujikura_wigigDevice}. Nonetheless, problems still surround the stability of connections, particularly with moving targets \cite{song_wigigComm}. For starters, the issue with obstacles and penetrations persists, and non-line-of-sight transmission without a relay of some sort is nigh impossible. Additionally, even variations of a 1 dBm in the signal strength can significantly decrease the achievable throughput \cite{fujikura_wigigDevice}. To maintain a network connection and take advantage of the highest throughput possible, multiple WiGig APs are needed with frequent handovers between them. However, handovers come with interruption time that cuts off the connection for a short interval, degrading the quality of experience \cite{qi_proactiveHandover}. This creates a complicated dilemma and tradeoff, where handovers interrupt and de-stabilize service for the users, but avoiding them means users are in a less-than-optimal connection and may lose network access altogether. \par

It is in this context that we will propose a proactive handover scheme for stable and efficient connection in a WiGig environment. Our framework is based on utilizing deep learning in the format of Convolutional Neural Networks (CNN) to predict the behavior of a user and their device. Then, based on this information, handover is performed in advance, before the signal degrades or is lost. Moreover, the behavior prediction allows our system to switch the connection to an AP that provides a stable environment in the future, further decreasing the need for handovers. \par

Our main contributions are listed as follows:

\begin{itemize}
\item{We provide a framework for collecting user data and training a CNN model capable of identifying patterns and predicting the future behavior of user devices in a WiGig environment.}
\item{We propose a novel handover decision scheme based on the predicted information that chooses APs with stable and efficient connections for each user device.}
\item{We provide simulation results based on experimental measurements that attest to how our proposal is capable of increasing the provided transmission rate and decreasing the number of handovers needed.}
\end{itemize}

The rest of this article is organized as follows. In the next section, we review what are the existing works on network prediction and handling handovers. After that, we present our proposed architecture to provide users with WiGig network access while simultaneously collecting the data needed to improve system efficiency. Following that, we will propose a framework that analyzes the collected data and smartly performs handovers. Finally, we evaluate the performance of our proposal and discuss some promising future directions while concluding the article.

\section{Proactive Handover and Network Prediction} 

Conventionally, networks and research on computer communications consider a reactive approach to handovers, where the connection from the user to the network is only changed to a new AP after the service is detected to have deteriorated \cite{yajnanarayana_proactiveHandover}. A big problem with this is that it forces the user to remain in a low-quality connection until the handover is completed. Such an issue is not ignored in the literature. Existing works have pointed out that networks with mobile users lead to frequent handovers and that using a deterministic and reactive scheme to control such handovers causes poor performance \cite{peng_proactiveHandover}. Especially in dense networks (which can be seen in WiGig, since the AP reach is easily blocked), a lot of redundant handovers (where a user is switched back and forth between APs) can happen, leading to high signaling overhead, handover latency, and service interruption, lowering the stability of the service \cite{hu_proactiveHandover}. Current literature (e.g., \cite{yajnanarayana_proactiveHandover}, \cite{peng_proactiveHandover}, and \cite{huang_proactiveHandover}) addresses this by using a proactive handover based on predicting future network states through Machine Learning (ML). However, this is often done while ignoring the application profile and requirements of the users \cite{qi_proactiveHandover}.  \par

The common ground of proactive handover research is predicting future network states. This has been done through user mobility and data transmission behavior, which are highly predictable as users commonly follow a limited number of patterns \cite{peng_proactiveHandover, huang_proactiveHandover}. This has also been done by predicting signal strength, which is tied strongly with user mobility \cite{qi_proactiveHandover}. Luckily for this field, the prediction of network states is widely studied, even unrelated to handover. The method is often the collection of past user data (packets generated, location, speed, signal strength, etc.) and identifying patterns for predicting future actions \cite{sepasgozar_networkPrediction, sakib_networkPrediction}. Additionally, the literature concludes that ML is the optimum method for making these predictions \cite{sepasgozar_networkPrediction, xiao_networkPrediction}. The ML models are trained on past user data and then used for predicting the future behavior of new users, based on the assumption that new users follow similar behaviors as past ones \cite{huang_proactiveHandover}. Moreover, the ML models are also updated during live use of the system, through reinforcement learning, which allows the models to stay updated even if the environment changes and continuously improve their accuracy \cite{rodrigues_smartNetwork}. Particularly, CNN has been identified as especially effective in finding patterns in network states and predicting future network parameters \cite{sakib_networkPrediction}. It is notable though that many of these works (e.g. \cite{sepasgozar_networkPrediction} and \cite{xiao_networkPrediction}) focus on just providing better predictions, without proposing any systems that can work on these predictions to improve practical network performance. \par

In this work, we address the shortcomings of the literature by providing a proactive handover scheme for WiGig networks. Our proposed system not only provides a method to collect user data, analyze it for patterns, and predict future network states but also how to use this prediction to provide practical performance improvement through better overall throughput. This is done while taking into consideration signal strength and also traffic generated by user applications. Prediction is done using CNN, following literature recommendation \cite{sakib_networkPrediction}. Lastly, as far as we know, this is the first study on network prediction-based handover management specifically for WiGig networks. This is important as WiGig has such high transmission and obstruction rates that conventional solutions should not be applied without careful investigation and adjustments \cite{song_wigigComm, bi_highFrequencyComm}.

\section{WiGig Environment and Data Collection} 

In this section, we will explain what is our assumed scenario that will outline the design and implementation of our proposed framework and handover scheme. Consider a room (or any area where the network is to be established) with multiple APs, all capable of offering WiGig access. The room also has what we will define as "points of interest" or PoIs. A PoI is somewhere where users are likely to stop and stand still for some time. For example, a bench in a park, a table in a restaurant, and a cashier at a store are all possible PoIs since users will probably walk toward those places and stop moving for a while. \par

Users will enter this room following a Gaussian distribution with a pre-determined average rate. Each user that enters the room will have a set of PoIs it wants to "visit" before leaving, and thus will move to each of those PoIs in sequence. Additionally, once they reach a PoI, they will stand still in that place for an amount of time, where the actual amount is influenced by the PoI (some PoIs, like a cashier line, have short staying time, while others, like a table in a restaurant, have long staying time). Finally, each user is using an application that fits into a pre-determined list of application types \cite{guo_6Gapplications}. Think of these types as video streaming, messaging, gaming, etc. Each type has its own, particular profile of data downloaded, data uploaded, and connection time. Additionally, we will also assume that some users are not compatible with some PoIs due to the application they are using and thus would not include those PoIs in their set of places to "visit". For example, a user watching a video or taking part in a video conference will probably not stand in a cashier line. \par

The scenario described is not only a realistic one but also provides a believable source of user behavior patterns. Since applications have particular profiles of data transmission and connection time \cite{qi_proactiveHandover}, analysis of these features makes it possible to identify the type of each user. Additionally, because there are a limited number of PoIs, there is also a limited number of paths between them. We can use user data to infer the location of the user as it moves and identify the path being taken \cite{huang_proactiveHandover}. Moreover, by knowing at which PoI the user is through this location information, it is possible to estimate how long the user will stay there due to the nature of that PoI. Finally, because each application type has a subset of compatible PoIs, the number of paths possible is also similarly further limited by the application being used, fortunately decreasing the number of possible patterns. \par

To finalize the assumed network system, we will consider that there is a server of some sort connected to all APs \cite{shen_networkConnectionAssociation}. At periodical intervals, every WiGig AP in the room will collect the following information: the signal strength between it and each user in the room, and how much data users connected to it have downloaded/uploaded in this time slot. The amount of transmitted data can be logged easily during communication with users. The signal strength does require an extra step, but it is not a troublesome one: the APs can broadcast a small message to all users that require a short acknowledgment, and this acknowledgment will be used to measure the signal strength (in fact, IEEE 802.11ay already has a similar mechanism implemented for beamforming training that can be adapted for this use). Note that the amount of transmitted data amount is logged only for connected users, while signal strength is logged for all users. This is done because the signal strength is useful for determining the estimated location of the user and having multiple data points allows for a rough triangulation \cite{sakib_networkPrediction}. This is not exact, but an approximate position should be useful in its own merit. All this collected information is sent to the server, which will aggregate it in a single tuple for each user for each time slot that contains: the amount of data downloaded/uploaded, and the signal strength between the user and each AP.

\section{Proposed WiGig Handover Scheme} 

\begin{figure}[!t]
\centering
\includegraphics[width=\linewidth]{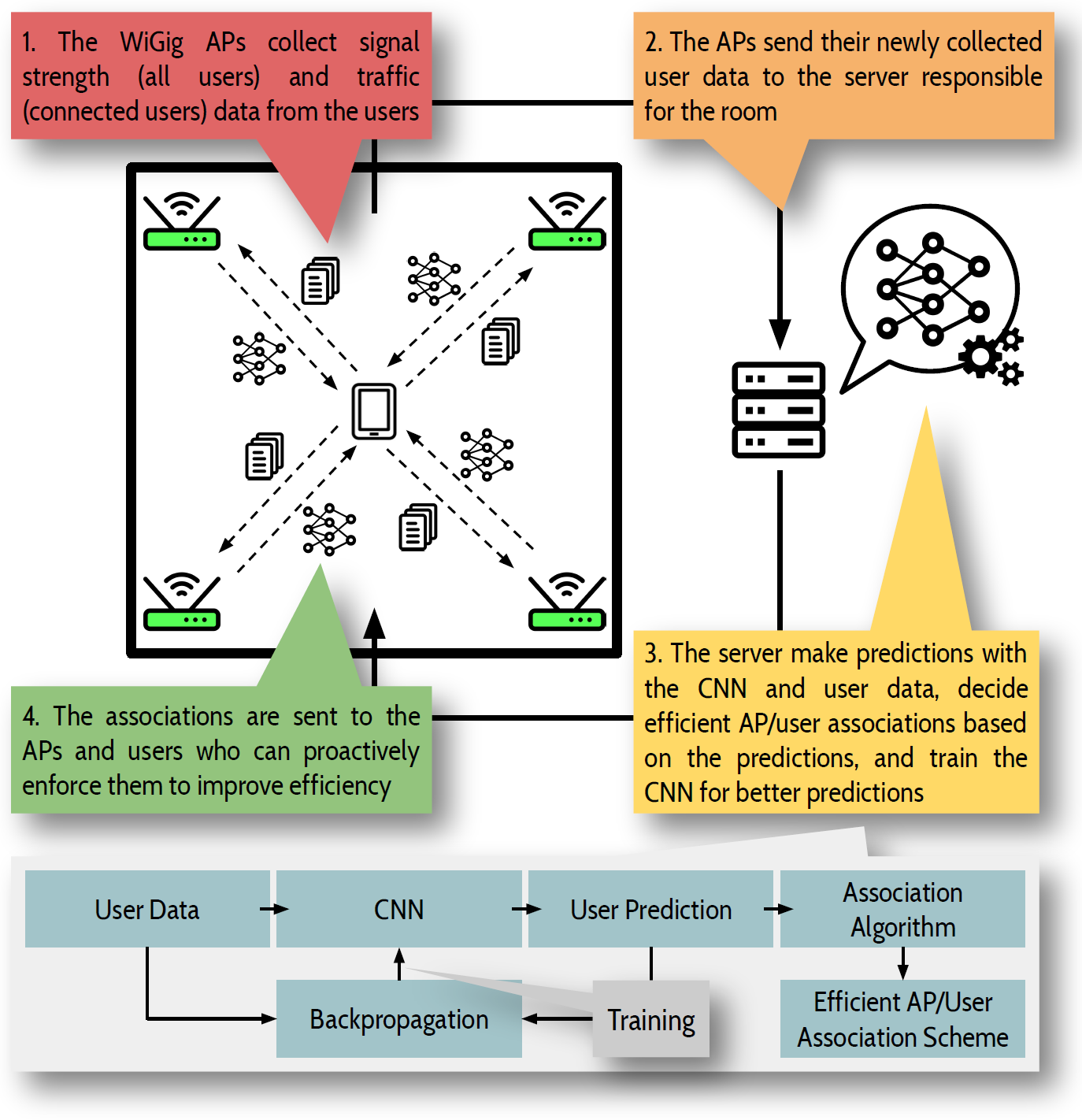}
\caption{The main loop of the proposed framework, showing how data is obtained to train the ML model, and how decisions based on the model are sent back to the agents.}
\label{proposal}
\end{figure}

By accumulating and logging user information in our server, we have created a suitable environment for learning-based analysis of the data collected from our network. The service model is shown in \figurename~\ref{proposal}, where we show two APs collecting data from one user and sending that data to the server. The server will use these data to train an ML model which is then used to predict future user behavior. This prediction is used for deciding how to realize handovers for the user, with the handover scheme being relayed back to the APs and the user so they can enact it. The chosen model for our proposed system will be a CNN due to its proven capability in predicting user behavior in networks and communication systems. CNNs have mechanisms called convolution and pooling that allow the learning model to intelligently select which features are more relevant from the input and use those to generate the desired output. In simple terms, CNNs can, better than human operators, identify which information is important to consider when trying to predict future user actions \cite{sakib_networkPrediction}. This information is fed into a regular neural network for generating the output desired. The output generated by the CNN is compared with what would be the correct output for the provided input and the difference between the values is used to update the weights of the whole model (neural network, pooling, and convolution sections) to produce more accurate outputs in the future. \par

In more specific terms, in our proposed framework, at each time slot, the server will, for each user, collect the most recent $X$ tuples of information to serve as input to the CNN. The output of the CNN will be the predicted information for the next $Y$ tuples for that particular user. I.e., the CNN will output what it predicts will be the future signal strength between the user and all APs as well as the transmission data for that user in the next future $Y$ time slots. The reason such prediction is possible is that there is a limited number of likely walking patterns that each user can take \cite{huang_proactiveHandover}. By looking at a big enough sample $X$ of historic data for that user, the pattern of information can be matched to one of the existing patterns, thus allowing for predicting the future behavior of that user. The way the model will learn of these patterns is through live training and reinforcement learning. Weights are set randomly at first. Then, as users connect to the system and information is collected, the model will make predictions. Because information continues to be collected, past predictions can eventually be compared with real collected data. For example, a prediction made at time slot $i$ will predict data up until time slot $i+Y$. Then, if the system collects data from the user up until time slot $i+Y$, the prediction made at $i$ can be compared with the corresponding "correct answer." This difference between the prediction and the correct result will be the basis for backpropagation that adjusts the weights \cite{rodrigues_smartNetwork}. Repeat the process enough times, with data from enough users, and the weights will be set so that patterns in that room can be identified. \par

The values for $X$ and $Y$ need to be chosen carefully. If they are too big, then the complexity of the CNN model will be big and may cause too much of a computational overhead to the system, which is a problem as it may delay handovers and defeat its initial purpose. Additionally, if $Y$ is too large, then it will be difficult to make accurate predictions, as predicting farther into the future is a more daunting task since there are more possibilities for the user to make choices to alter the pattern it fits into. However, if $Y$ is too small, then there is little benefit to be gained, as we are not seeing enough of the user's future behavior to make the best decisions for a stable association. Moreover, if $X$ is too small, then we risk not providing enough input data to properly identify the user's pattern. So, while CNN is a proven adequate model for pattern prediction, it must be properly tuned, which often comes from empirical studies \cite{sakib_networkPrediction}. \par

It is worth pointing out that the ML model has no prior knowledge about the scenario. I.e., there is no need to have pre-determined data regarding the location of PoIs, details of the applications, which application each user uses, etc. The CNN will detect these features live, as the network is used and data is collected. This results in a feasible deployment plan, as the system can just be plugged in and learn the patterns by itself \cite{wang_quantumSmartNetworking}. Moreover, the server mentioned here just needs enough processing capabilities to execute one forward run and one backpropagation of the CNN for each currently connected user per timeslot. If there are N users and the CNN has M values to process, the total complexity is $O(NM)$. Depending on the size of the room and CNN, this can be done with a small-scale server or even one of the APs standing in as a server. \par

\begin{algorithm}[!t]
\caption{AP association based on prediction.}\label{alg:alg1}
\begin{algorithmic}
\STATE 
\STATE \textrm{start with all APs having $0$ associated users}
\STATE {\textbf{for}} \textrm{user} $i$ \textrm{from} $0$ \textrm{to} $N-1$
\STATE \hspace{0.5cm} {\textbf{for}} \textrm{AP} $j$ \textrm{from} $0$ \textrm{to} $P-1$
\STATE \hspace{1.0cm} \textrm{calculate throughput of $i$ with $j$ based on prediction}
\STATE \hspace{1.0cm} \textrm{update max throughput appropriately}
\STATE \hspace{0.5cm} {\textbf{if}} \textrm{max throughput $-$ current throughput $>$ threshold}
\STATE \hspace{1.0cm} \textrm{associate $i$ with AP with max throughput}
\STATE \hspace{0.5cm} {\textbf{else}}
\STATE \hspace{1.0cm} \textrm{associate $i$ with current AP}
\end{algorithmic}
\label{auxiliaryAlgorithm}
\end{algorithm}

Finally, it is important to explain how the predicted data is used for deciding user/AP association and handover. In our system, the output of the CNN for all users is fed into an auxiliary algorithm. This algorithm, shown in Algorithm \ref{auxiliaryAlgorithm} (where $P$ is the number of APs), receives as input the predicted signal strength and transmitted data for each user and the current AP association of each user, and greedily checks all APs for the one that offers the highest transmission rate in the future $Y$ time slots. This is done iteratively, going through each user and associating it with the best AP available based on the user's future behavior. As APs get connected to more users, the actual throughput that can be provided to each user decreases (since access is provided in a time-division way in WiGig, more connected users mean less access time per user, which leads to lower rates \cite{bi_highFrequencyComm}). The algorithm will take this, the number of connected users, alongside the predicted future behavior into account when choosing the AP that offers the highest rate for each user \cite{shen_networkConnectionAssociation}. Lastly, the algorithm will also refrain from making users change APs if the resulting improvement in transmission rate is not significant. Overall, considering the execution and training of the CNN, the whole scheme has a complexity of $O(NM + NYP)$.

\section{Performance Evaluation} 

\begin{table}[!t]
\caption{Parameters Used in The Performance Evaluation}
\label{parameterTable}
\centering
\begin{tabular}{|c||c|}
\hline
\textbf{Parameter Name}				 	& \textbf{Parameter Value}\\ \hline
Room Size							 	& 300m x 300m\\ \hline
Time Slot Length						 	& 1s\\ \hline
Number of APs				 			& 4\\ \hline
Number of Input Time Slots	 			& 25\\ \hline
Number of Output Time Slots	 			& 10\\ \hline
User Interarrival Time Rate	 			& 10s\\ \hline
Number of Application Types	 			& 3\\ \hline
User Data Generation			 			& 10 - 1000Mbps\\ \hline
User Movement Speed			 			& 0.1 - 2.0m/s\\ \hline
Total Number of PoIs						& 4\\ \hline
PoI Stay Time				 			& 1 - 100s\\ \hline
Number of PoIs per User		 			& 1 - 3\\ \hline
Throughput Threshold for Handover	 	& 200Mbps\\ \hline
Achieved Throughput Rate				 	& From IEEE 801.11ay \cite{fujikura_wigigDevice}\\ \hline
\end{tabular}
\end{table}

Simulation tests were carried out to evaluate the performance of the proposed solution. 10000 simulation runs were done, with the random deployment of APs and PoIs, and the results shown are the average across all runs. The users' features, such as which application they use or which PoIs they visit, are also determined randomly. Which PoIs are available for each application is also determined randomly. This is all done to achieve enough statistical relevance for our results. Finally, the signal strength between users and APs is determined by distance and was derived based on real-life measurements performed using a "Fujikura 60GHz mmWave Wireless Communications Module" \cite{fujikura_wigigDevice}. These experiments measured the signal strength while varying the distance between the two devices. The values obtained were used in the simulation to add more realism to the results seen in this paper. Unless stated otherwise, the parameters used for all graphs shown here are in Table \ref{parameterTable}, which were obtained from the literature \cite{song_wigigComm, bi_highFrequencyComm}. The CNN hyperparameters come from extensive empirical studies looking for the best performance, omitted here for brevity. \par

\begin{figure}[!t]
\centering
\includegraphics[width=\linewidth]{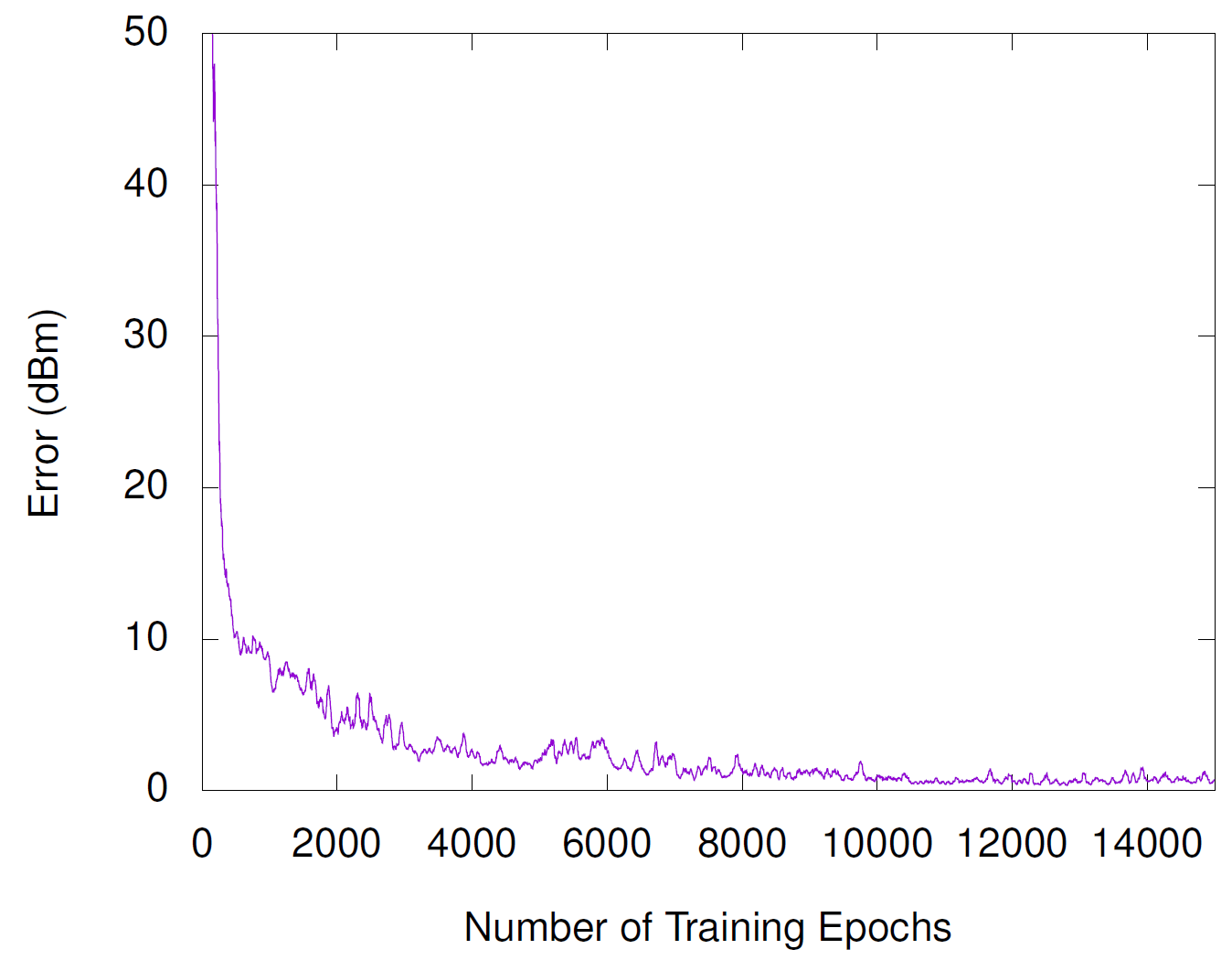}
\caption{How the model improves in its prediction accuracy as it is allowed to train and see more data from the environment.}
\label{predictionAccuracy}
\end{figure}

First, we measured what is the error value given by the CNN when predicting signal strength between a user and the APs. Signal strength is very important for determining transmission rate and estimating user location, plus it can vary significantly while the user is moving, which is why it was selected to highlight the prediction performance \cite{sepasgozar_networkPrediction}. Epoch here is determined by one time slot, where on each time slot, the most recent 25 tuples of each user are used as input to determine the next 10 tuples. Results can be seen in \figurename~\ref{predictionAccuracy}. As expected, the error is high at the beginning since the CNN is predicting randomly without any a priori learning. However, this quickly changes. After 5000 epochs, the average error is below 5 dBm. After 10000 epochs, the error is below 1 dBm. This gives us two insights. First, it is better to not use the AP association generated from the CNN predictions until learning reaches an acceptable level. Second, the CNN is definitely capable of learning the patterns of any room that follows our assumed scenario and predicting future user behavior with minimal error. \par

\begin{figure}[!t]
\centering
\includegraphics[width=\linewidth]{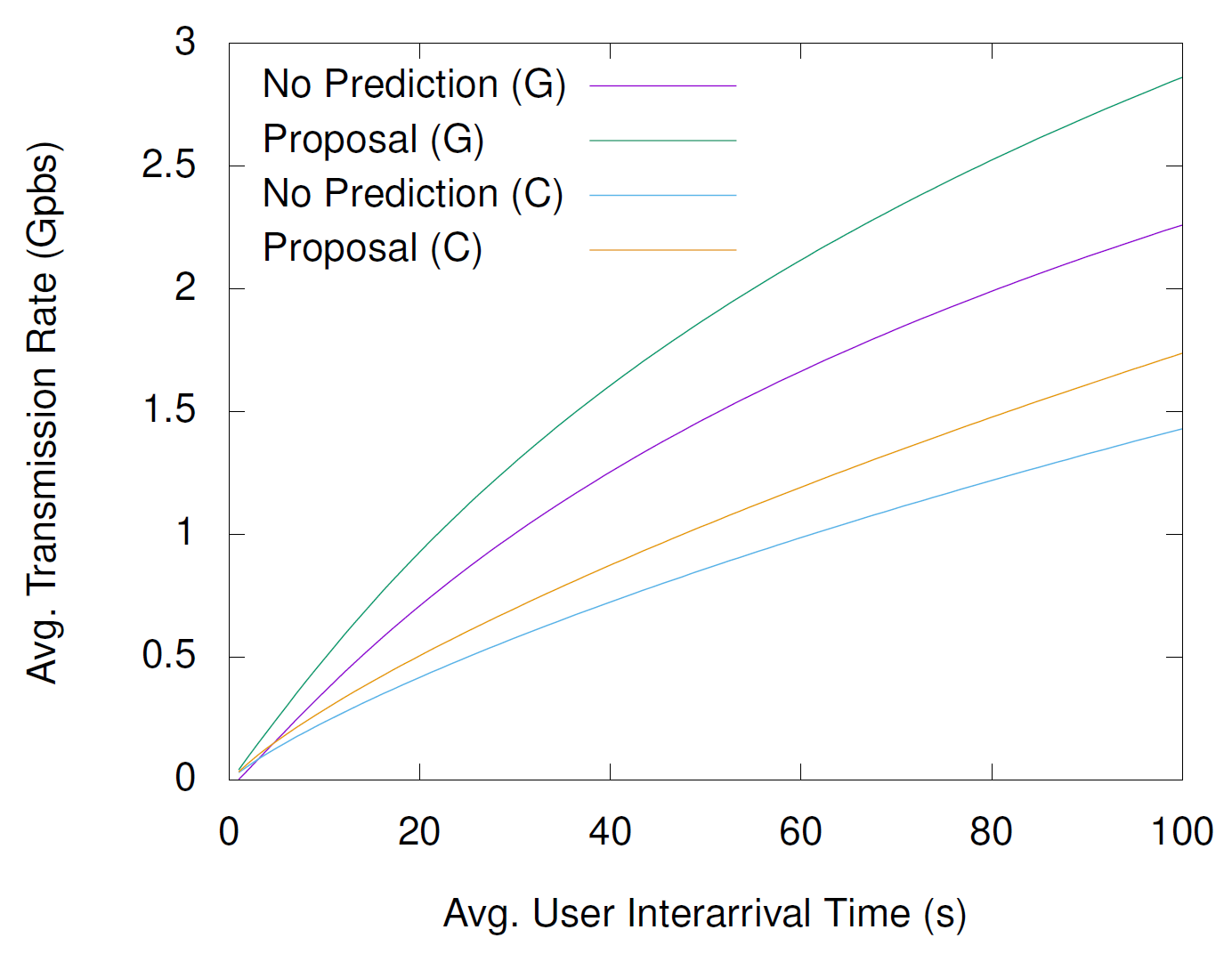}
\caption{What is the transmission rate provided by all schemes for different values of network workload represented by user interarrival time.}
\label{interarrivalTime}
\end{figure}

\begin{figure}[!t]
\centering
\includegraphics[width=\linewidth]{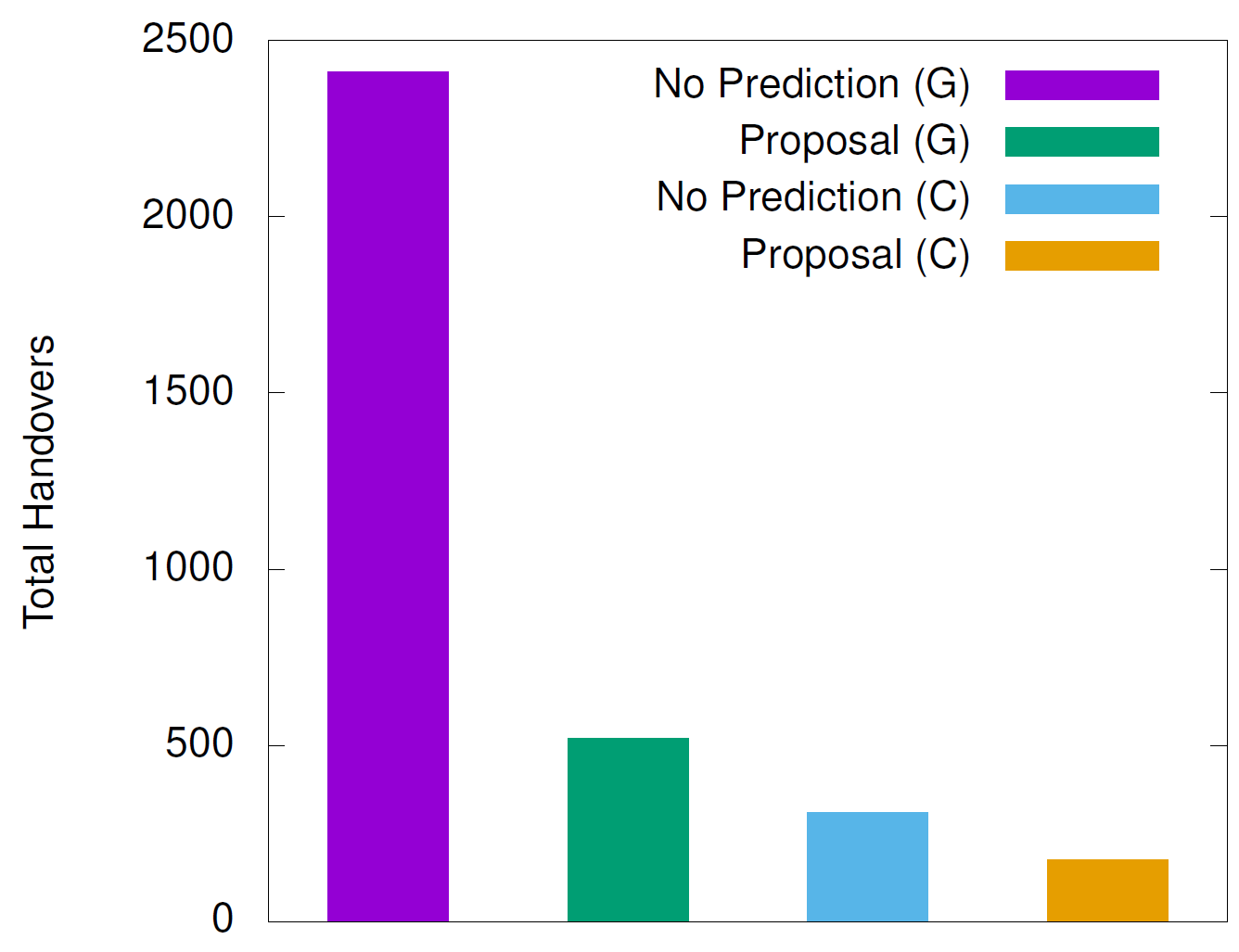}
\caption{The number of handovers that were performed with each method.}
\label{totalHandovers}
\end{figure}

Next, we measured what is the average transmission rate achieved by the proposal. For comparison, 3 other methods are presented. First, we have a No Prediction method, where the system does not attempt to predict future user behavior. Instead, user/AP association is decided at each time slot based solely on the current signal strength and maximizing current achievable transmission. In contrast, the proposal tries to maximize the achievable transmission rate in the future 10 time slots instead of the current one only. Additionally, for both the proposal and the No Prediction method, two variations were tested. The greedy variation (represented by G in the graphs) will always perform handovers if a better transmission can be achieved, regardless of whether the improvement is big or not. The conservative variant (represented by C in the graphs), on the other hand, avoids making handovers unless the transmission rate can be improved by at least 200 Mbps. Results are shown in \figurename~\ref{interarrivalTime}, where we varied the user interarrival time (high interarrival values mean more time between the arrival of each user and thus fewer users in the system). As expected, fewer users mean less competition for network access and higher achievable transmission rates. The conservative variants offer lower rates than their greedy counterparts. This shows that, in WiGig, switching to the best AP, to some extent, offers better performance as the system keeps trying to optimize the offered service. However, some stability is needed, and this is illustrated by how the proposal outperforms the No Prediction method in both variants, offering upwards of 1 Gbps extra. To explain why, we calculated how many handovers are done in each method, shown in \figurename~\ref{totalHandovers}. As expected, the conservative solutions do fewer handovers, since there is a more strict trigger for changing APs. Moreover, the No Prediction method has more handovers as it makes no effort to look for a stable connection (just the best one at the moment). Meanwhile, both figures tell us that the proposal not only is effective in finding APs that demand fewer handovers and offer a more stable environment (a reflection of how the CNN is capable of predicting future user actions) but also that having fewer handovers does have a significant impact in the performance of the system. Moreover, the conservative methods seem to lean too heavily toward avoiding handovers and the transmission rates suffer as a result, behaving even worse than the greedy No Prediction solution despite the high number of handovers. This points toward a careful tradeoff between handovers and transmission rate optimization. Nonetheless, it is clear from all results that the proposal is the best solution for predicting user behavior and finding optimal AP association.

\section{Conclusions} 

In this paper, we presented a new framework aimed toward WiGig systems that improve the average transmission rate through smart prediction of user behavior. A CNN was utilized for learning user patterns in a room and then predicting channel data surrounding a user in the form of signal strength in relation to APs and transmitted data. This prediction was used for choosing user/AP associations in a way that the transmission rate is improved while handovers are avoided and users are presented with stable connections. Additionally, by predicting future user behavior, handovers could be done proactively, before the connection degraded. Performance evaluation showed that not only the CNN effectively predicts user behavior, but the proposed algorithm based on this prediction is successful in improving the transmission rate and avoiding unnecessary handovers. In the future, we would like to study how to effectively handle  situations where the room layout changes and scenarios where line-of-sight with the AP is not guaranteed.

\section*{Acknowledgments}
These research results were obtained from the commissioned research (No. 22403) by the National Institute of Information and Communications Technology (NICT), Japan.

\bibliographystyle{IEEEtran}
\bibliography{Reference}


\section*{Biographies}

\begin{IEEEbiographynophoto}{Tiago Koketsu Rodrigues}
[M'15], previously Tiago Gama Rodrigues, is currently an Assistant Professor at Tohoku University. His research interests include artificial intelligence, machine learning, network modeling and simulation, and cloud systems. From 2017 to 2020, he was the Lead System Administrator of the IEEE Transactions on Vehicular Technology, overviewing the review process of all submissions and the submission system as a whole. He serves as an Editor for the IEEE Transactions on Vehicular Technology and the IEEE Network.
\end{IEEEbiographynophoto}

\begin{IEEEbiographynophoto}{Shikhar Verma}
[M'21] received the bachelor’s degree in computer science and engineering from the National Institute of Science and Technology, Berhampur, India, in 2014, and the M.Sc. and Ph.D. degrees from the Graduate School of Information Sciences (GSIS), Tohoku University, Sendai, Japan, in 2018 and 2021, respectively. Since 2021, he has been a Research Assistant Professor with GSIS, Tohoku University. Dr. Verma was also a recipient of the prestigious MEXT Scholarship and JSPS Fellowship. He also received the Dean's Award from Tohoku University in 2021 and the Best Paper Award at IEEE ICC in 2018.
\end{IEEEbiographynophoto}

\begin{IEEEbiographynophoto}{Yuichi Kawamoto}
[M'17] received the M.S. and Ph.D. degrees in information science from Tohoku University, Sendai, Japan., in 2013 and 2016, respectively. He is currently an Associate Professor with the Graduate School of Information Sciences, Tohoku University. He has authoured or coauthored more than 60 peer-reviewed papers, including several high-quality publications in prestigious IEEE journals and conferences. His research interests include satellite communications, unmanned aircraft system networks, wireless and mobile networks, ad hoc and sensor networks, green networking, and network security. Dr. Kawamoto was the recipient of the Best Paper Awards witht many international conferences, including IEEE flagship events, such as the IEEE Global Communications Conference in 2013, IEEE Wireless Communications and Networking Conference in 2014, and IEEE International Conference on Communications in 2018. He was also the recipient of the prestigious Dean's Award and President's Award from Tohoku University in 2016. He is a Member of the Institute of Electronics, Information, and Communication Engineers.
\end{IEEEbiographynophoto}

\begin{IEEEbiographynophoto}{Nei Kato}
[M'04, SM'05, F'13] is a full professor and the Dean with Graduate School of Information Sciences, Tohoku University. He has researched on computer networking, wireless mobile communications, satellite communications, ad hoc \& sensor \& mesh networks, UAV networks, smart grid, AI, IoT, Big Data, and pattern recognition. He is the Editor-inChief of IEEE Internet of Things Journal. He has published more than 500 papers in prestigious peerreviewed journals and conferences. He served as the Vice-President (Member \& Global Activities) of IEEE Communications Society (2018-2021), and the Editor-in-Chief of IEEE Transactions on Vehicular Technology (2017-2021). He is a fellow of The Engineering Academy of Japan, a Fellow of IEEE, and a Fellow of IEICE.
\end{IEEEbiographynophoto}

\begin{IEEEbiographynophoto}{Mostafa M. Fouda}
[M'11, SM'14] is an Assistant Professor with the Department of Electrical and Computer Engineering at Idaho State University, ID, USA. He also holds the position of Full Professor at Benha University, Egypt. He received his Ph.D. in Information Sciences from Tohoku University, Japan, in 2011. His research interests include cyber security, machine learning, IoT, and 6G networks. He has served on the technical committees of several IEEE conferences.  He is also a Reviewer in several IEEE Transactions and Magazines. He is an Editor of IEEE Transactions on Vehicular Technology (TVT) and an Associate Editor of IEEE Access.
\end{IEEEbiographynophoto}

\begin{IEEEbiographynophoto}{Muhammad Ismail}
[M'13, SM'16] is an Assistant Professor with the Department of Computer Science, Tennessee Technological University, Cookeville, TN, USA. He received his Ph.D. in Electrical and Computer Engineering from the University of Waterloo, Canada, in 2013. His research interests include machine learning, 5G+ wireless networks, and cyber-physical security. He is a co-recipient of six best paper awards from IEEE conferences. He has served on the technical committees of several IEEE conferences. He is an Editor of IEEE Transactions on Vehicular Technology (TVT) and IEEE Internet-of-Things (IoT) Journal.
\end{IEEEbiographynophoto}

\vfill

\end{document}